\documentclass[a4paper]{uniba}
\usepackage{amssymb}
\usepackage{graphicx}

\newcommand{\keywords}[1]{\par\addvspace\baselineskip
\noindent\keywordname\enspace\ignorespaces#1}

\usepackage[english]{babel}
\usepackage{url}

\usepackage{metalogo}

\usepackage{cite}
\begin{document}

\mainmatter  

\title{Problem Complexity in Parallel Problem Solving}

\titlerunning{Problem Complexity in PPS}

%
%
\author{Sebastian Herrmann%
\thanks{Corresponding author}, %
 J\"orn Grahl %
\and Franz Rothlauf}
\authorrunning{Problem Complexity in Parallel Problem Solving}

\institute{Dept. of Information Systems and Business Administration,\\
Johannes Gutenberg-Universit\"at,\\
Jakob Welder-Weg 9, 55128 Mainz, Germany\\
\url{http://wi.bwl.uni-mainz.de}\\
\url{{s.herrmann, grahl, rothlauf}@uni-mainz.de}}

%
%

\toctitle{Problem Complexity in Parallel Problem Solving}
\tocauthor{Problem Complexity in Parallel Problem Solving}
\maketitle

\begin{abstract}

Recent works examine the relationship between the communication structure and the performance of a group in a problem solving task. Some conclude that inefficient communication networks with long paths outperform efficient networks on the long run. Others find no influence of the network topology on group performance. We contribute to this discussion by examining the role of problem complexity. In particular, we study whether and how the complexity of the problem at hand moderates the influence of the communication network on group performance. Results obtained from multi-agent modelling suggest that problem complexity indeed has an influence. We observe an influence of the network only for problems of moderate difficulty. For easier or harder problems, the influence of network topology becomes weaker or irrelevant, which offers a possible explanation for inconsistencies in the literature.  

\end{abstract}

\section{Problem Solving in Groups}
Humans routinely assemble into groups to solve complex problems. As they exchange ideas and approaches, the performance of the group in solving a problem depends on the individual performance of the group members and on the type and structure of the collaboration. A question that has widely been discussed in literature is how the communication network's structural properties influences group performance. The topic is of interest to psychology, sociology as well as management and organization science.
Some studies find that complete freedom of communication can be more limiting than restricted communication patterns\cite{Leavitt.1951, Guetzkow.1955, Cohen.1962}. Other studies come to conflicting conclusions or a more differentiated view where the optimal group structure depends on the tasks applied in the experiments \cite{Shaw.1954, Mulder.1960, Carzo.1963}. Most studies draw conclusions from small group experiments or observational data. Recent research in the computational social sciences tries to overcome this limitation by running multi-agent models or web-based experiments with large groups \cite{Lazer.2009}.

Lazer~\& Friedman \cite{Lazer.2007} try to identify superior network structures in a scenario called \emph{parallel problem solving}. Here a set of roughly equivalent actors attempts to solve a complex problem. They conduct an agent-based simulation (hereafter referred to as \emph{LF-model}). The agents search for good solutions in an $NK$-landscape~\cite{Kauffman.1989}, a well-established model for problem representation in organizational theory \cite{Levinthal.1997}. 
A solution is represented as a binary string of $N$ bits and has a score that agents try to maximize. They can \emph{explore} from a given solution to another by flipping a random bit. Alternatively, they can \emph{exploit} solutions of their network neighbors by copying them if they are better. The problem space is multi-modal and moderately rugged. Search can lead to a state in which no further improvement is achievable even though that state is not the global optimum. Lazer~\& Friedman average the scores of all agents to measure the performance of the group. They compare the group performance for several networks. The networks are characterized by differences in efficiency (their average path length). Networks with higher efficiency (smaller path lengths) are able to disseminate information faster than less efficient networks. Their results suggest that efficiency is beneficial for short-run performance. On the long run however, less efficient networks perform better.

Mason~\& Watts~\cite{Mason.2012} conduct a series of web-based experiments with human subjects (hereafter referred to as \emph{MW-model}). The subjects play a networked game with the objective to select points for oil-drilling on a map. They do not know where the good and where the bad oil wells are positioned on the map, but they are able to see the coordinates and earnings of their network neighbors. Contrary to \cite{Lazer.2007} they find that the average path length of the network that connects the subject is negatively correlated with the mean earnings achieved by the group. Put differently, networks with shorter paths perform better. They do not find a significant difference in the probability to find the best solution, even though efficient networks have a slightly higher performance.

Lazer~\& Friedman conclude that an efficient network negatively affects information diversity which in turn negatively affects group performance. Higher network efficiency leads to lower average scores. Mason~\& Watts find no significant difference between the networks in terms of success probability. As efficient networks have higher earnings on average, they presume that---in case of doubt---efficient information flow can only be advantageous. We believe that both models provide valuable insights into basic mechanisms in collaboration and indicate that communication structure is important. Nevertheless, it is an open question whether the results are substantially conflicting or if they are caused by different assumptions or experimental conditions. 

We believe that a deeper understanding of the network influence on group performance requires that additional factors are included into the analysis. A glimpse into small groups studies reveals that several other factors have already been discussed. Heise~\& Miller \cite{Heise.1951} find that for simple problems, groups with more communication channels produce less errors, for intermediate problems more tightly organized groups are better. For high complexity, there is no difference at all between groups. Others find the opposite~\cite{Shaw.1954b}: for simple problems, group structures makes no difference in accuracy, but for complex problems, less tightly structured groups produce less errors. Even though these studies have contradictory findings, both authors agree that there is an influence of the task complexity. Also Lazer~\& Friedman propose that complexity should be taken into consideration in their model. Mason~\cite{Mason.2013} suspects that task complexity (``the potential for local maxima'') could be a necessary condition for the superiority of networks with long paths in the long run. Hence, the goal of this paper is to provide first insights into how problem complexity moderates the network influence on group performance. 

\section{Experiments and Results}
In the original experiments on the LF-model the agents have to solve problems of moderate complexity. We conduct a series of experiments with a modified LF-model. Our intention is to test the impact of network structure on problem-solving capability for different levels of task complexity. A side condition is to make as little changes as possible to the original model, thus the group size of 100~agents per network, behavioral parameters etc. are left unchanged. Yet a presumption we have to reconsider is that of the operationalization for group performance. Comparing the original model to real-world challenges---e.g. research on composition of an effective pharmaceutical---we see a lack of construct validity: these problems are analogous to combinatorial optimization problems, where the detection of the optimal solution is desirable (so-called conjunctive tasks)~\cite{Steiner.1972}. Under this assumption, the average degree of target achievement of the group members (as assumed in \cite{Lazer.2007}) is rather inaccurate. Hence, we define group performance as the probability to find the optimal solution.

A rigorous approach to test the impact of network structure for a particular task should consider all networks that can be evolved for a given number of agents.
This is impracticable because the number of possible networks grows exponentially with the number of actors. We therefore measure network influence as the difference in long-term performance between a linear and a fully connected network. These structures have been studied in \cite{Lazer.2007} and they represent both extremes of the concept of average path length. A linear network has the highest average path lengths attainable with a given number of nodes, a totally connected one has the lowest average path lengths. 

A benefit of $NK$-landscapes is that their complexity can be tuned by the parameter $K$, which determines the number of interdependencies between the $N$ binary decision variables. A value of $K = 0$ results in a smooth uni-modal, easy-to-solve space, whereas $K = N - 1$ causes a maximally complex space. 
We test instances of the LF-model for a variety of difficulty levels. 
For this purpose 100 random $NK$-landscapes are generated for each $K \in [0,~19]$ ($N = 20$ bits). We execute 100 repetitions for each space and then count in how many cases at least one agent finds the optimal solution.

\begin{figure}
	\centering
	\includegraphics[height=6.2cm]{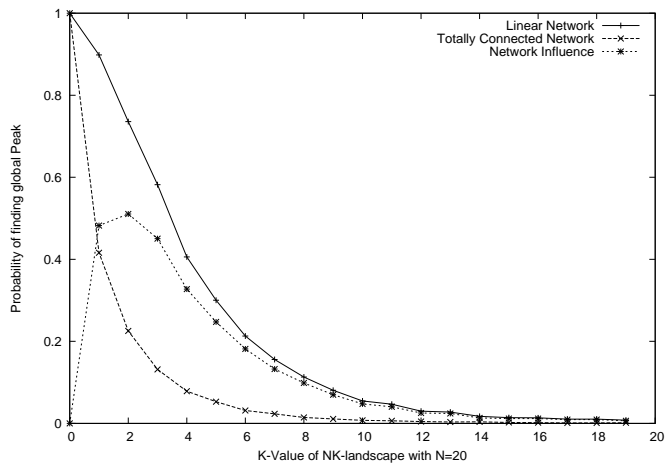}
	\caption{The vertical axis shows the probability of finding the optimal solution for linear and totally connected networks in 100 random $NK$-landscapes. The horizontal axis shows the parameter~$K$ as a measurement of problem complexity, from very easy ($K=0$) to very hard ($K=19$). Network influence is the difference between the success probabilities of the two networks. Lines between points are for illustration purposes only.}
	\label{fig_lf}
\end{figure}

Fig.~\ref{fig_lf} shows the result. For intermediate problem difficulty, the linear network outperforms the totally connected network on the long run, but is inferior on the short run. This is in accordance with the original model, even though our metric for group performance is different. However, the more we increase or decrease problem complexity from this point, the differences in performance between both networks vanish: the network influence falls towards zero as $K$ approaches $0$ or $19$. We observe a clear influence of network structure on the groups' success probability for intermediate problem difficulty only and reduced or even absent influence to both extreme sides of the complexity metric. Imagining a line between the data points for network influence in fig.~\ref{fig_lf}, a curved shape appears. 

\begin{figure}
	\centering
	\includegraphics[height=6.2cm]{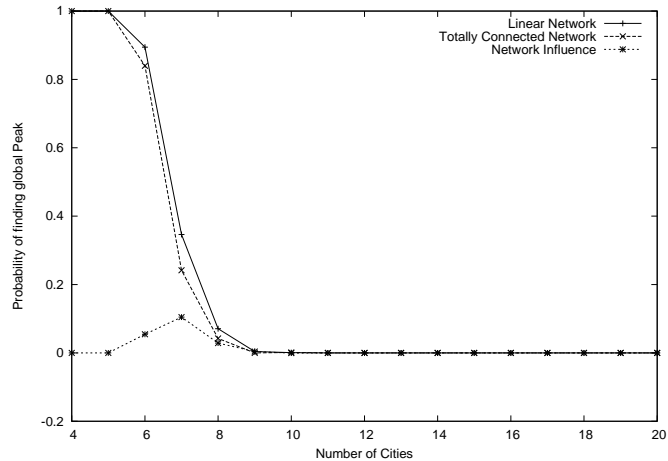}
	\caption{The vertical axis shows the probability of finding the optimal solution for linear and totally connected networks in 100 random TSP-instances. The horizontal axis represents the number of cities as the parameter for problem complexity. Network influence is defined as difference between the success probabilities of the two networks. Lines between points are for illustration purposes only.}
	\label{fig_tsp}
\end{figure}

To assure that the curvilinear relationship is not a peculiarity of the $NK$-landscape we let the agents solve a Traveling Salesman Problem~(TSP). The goal of a TSP is to connect places on a map in a single route so that the total distance for the entire tour is minimized. Using the TSP is a first step into the direction of understanding network influence on  more realistic settings. Full implementation details are in \cite{Lazer.2013}. In our case the complexity of the TSP is constant for a given number of cities \cite{Stadler.1996}. Hence, to vary the complexity we vary the number of cities from $1$ to $20$. The success of the group is again the probability that at least one member finds the optimal solution. The results confirm our previous finding~(fig.~\ref{fig_tsp}). We observe differences in performance only for intermediate problem complexity and diminishing effects for extreme difficulty levels. This suggests that the curvilinear relationship between problem complexity and network influence is not a peculiarity of the $NK$-landscape.

\section{Summary and Conclusion}
Our analyses suggest that the influence of network structure on group performance is more nuanced than previously thought. There is high evidence for a curvilinear moderating influence of task difficulty on the relationship between a group's network structure and its performance. People may be able to solve very easy problems, independent from their structure of collaboration. For very hard problems, collaboration in any form may have no effect either as the problem is so unstructured that the exploitation of promising ideas from others will not lead actors towards a direction in which the optimal solution can be found. For intermediate problems, the network structure affects a groups ability to find appropriate solutions and an adequate balance of exploration and exploitation is necessary. 
Our findings also indicate that recent results on the influence of network structure may not be as contradictory as they appear. Insignificant network effects could be an artifact of the task complexity. For instance, Mason~\& Watts~\cite{Mason.2012} found no significant difference between networks in terms of success probability. It could be possible that the problem used in their setup is too easy or too hard for a network influence to play a significant role. 

It is difficult to draw managerial conclusions from computational studies. Future research should try to replicate our findings with human subjects. 

\bibliographystyle{splncs}

\end{document}